\documentclass[11pt]{article}
\usepackage{latexsym,amsmath,chicago,fullpage,wrapfig,graphicx,epsfig}

\newcommand{\sbf}{\boldsymbol}
\newcommand{\sla}{{\scriptscriptstyle\langle}}
\newcommand{\sra}{{\scriptscriptstyle\rangle}}

\title{\bf Retrospective Statistical Inference}
\author{Chong Gu\\{\it Department of Statistics, Purdue University}}
\date{}

\begin{document}
\maketitle

\begin{abstract}
  In this article, we explore a new paradigm for statistical
  inference.  The approach centers around the point estimate based on
  observed data, simulating replicates using the estimate as the truth
  to produce clones of the estimate, with inference deriving from the
  clone distribution.  It avoids prospective finite-dimensional model
  assumptions, but it makes no probabilistic claims concerning the
  truth; it suggests an alternative system of uncertainty
  quantification that is operable in nonparametric function
  estimation.  The procedures are demonstrated using examples of
  smoothing spline ANOVA models in nonparametric regression.  The
  paradigm also applies in parametric regression, where the proposed
  inference closely resembles traditional inference operation-wise.
  Conceptual discussions are scattered throughout.

\end{abstract}

\section{Introduction}

Traditional statistical inference procedures, such as confidence
intervals and hypothesis testing, rely on sampling distributions of
sample statistics involved.  Sampling distributions are derived under
assumptions, among which an essential provision is some prospective
finite-dimensional model space.  In settings where finite-dimensional
model spaces can not be specified a priori, such as with nonparametric
regression, sampling distributions are not available, so traditional
inference is largely infeasible, but the need for systematic
uncertainty quantification remains in those settings.

As an attempt on alternative reasoning, we explore a different
paradigm for statistical inference.  We assume and rely on a
probabilistic data generation mechanism and a procedure to obtain a
point estimate, but we make no attempt on probabilistic claims
concerning the unknown truth.  Instead, the proposed retrospective
inference is anchored on the point estimate.

Specifically, using the point estimate as the truth, one may simulate
replicates according to the data generation mechanism, then follow the
``same procedure'' to obtain point estimates based on the replicates;
such estimates will be called clones of the original point estimate.
The proposed retrospective inference derives from the clone
distribution.  The delicate part of the exercise is the mapping from
the simulated replicates to the clones; in nonparametric settings, the
latter depend not only on the former but also on some tuning
parameter(s), of which the practical selection is essential to the
viability of the proposed paradigm.

Working from the known point estimate, one no longer needs prospective
characterizations of the unknown truth, though such information, if
available, does no harm.  In fact, in parametric regression, the clone
distribution is often analytically tractable to some extent, and
retrospective inference closely resembles traditional inference
operation-wise.

The rest of the article is organized as follows.  We shall demonstrate
the work flow in the settings of nonparametric regression using
smoothing spline ANOVA models, of which some background materials are
reviewed in Section~2.  Section~3 discusses smoothing parameter
selection, i.e., the clone definition, with emphasis on the role of
effective constraints; optimal performers are also suggested as a
possible fallback option in settings where effective constraints are
not available or quantifiable.  Confidence bands and hypothesis
testing based on clone distributions are demonstrated in Section~4
using a few simulated and real-data examples.  Section~5 puts
retrospective inference to work in standard parametric regression,
showing its close resemblance to traditional inference.  Further
conceptual discussions are collected in Section~6 to conclude the
article.

\section{Regression using Smoothing Spline ANOVA Models}

We shall demonstrate retrospective inference in the settings of
nonparametric regression using smoothing splines, of which some
background materials are reviewed here.

\subsection{Cubic Smoothing Spline and Penalized Likelihood}

Consider $Y_{i}=\eta(x_{i})+\epsilon_{i}$, $i=1,\dots,n$, where
$x_{i}\in[0,1]$, $\epsilon_{i}\sim{N}(0,\sigma^{2})$ independent.  One
may estimate $\eta(x)$ via the minimization of
\begin{equation}\label{cubic}
  \frac{1}{n}\sum_{i=1}^{n}\big(Y_{i}-\eta(x_{i})\big)^{2}
  +\lambda\int_{0}^{1}\ddot{\eta}^{2}(x)dx,
\end{equation}
where $\ddot{\eta}=d^{2}\eta/dx^{2}$, $\lambda\in(0,\infty)$.  The
minimizer of (\ref{cubic}) is known as a cubic smoothing spline.

The above is a special case of the general penalized likelihood
method.  To estimate $\eta(x)$ on a generic domain $x\in\mathcal{X}$
using stochastic data, one may minimize
\begin{equation}\label{pelk}
  L(\eta|\text{data})+\lambda{J}(\eta),
\end{equation}
where $L(\eta|\text{data})$ is usually taken as the minus log
likelihood of the data, $J(\eta)$ is a quadratic roughness functional,
and the smoothing parameter $\lambda$ controls the trade-off between
the goodness-of-fit and the smoothness of the $\eta$ estimate.  In
(\ref{cubic}), $L(\eta|\text{data})$ is proportional to the minus log
likelihood of Gaussian responses and
$J(\eta)=\int_{0}^{1}\ddot{\eta}^{2}(x)dx$.

Observing $Y_{i}\sim\text{Bin}\big(m_{i},p(x_{i})\big)$ for
$x_{i}\in[0,1]$, one may perform logistic regression via the
minimization of
\begin{equation}\label{logit}
  -\frac{1}{n}\sum_{i=1}^{n}\big\{Y_{i}\eta(x_{i})-m_{i}\log(1+e^{\eta(x_{i})})\big\}
  +\lambda\int_{0}^{1}\ddot{\eta}^{2}(x)dx,,
\end{equation}
where $\eta=\log\frac{p}{1-p}$ is the logit.

Extensive discussions of penalized likelihood regression can be found
in \citeN[Chap.~3, Chap.~5, Sect.~8.6]{gu:13}.

\subsection{Reproducing Kernel Hilbert Spaces}\label{rkhs}

Technically, the minimization of (\ref{pelk}) takes place in the space
$\{\eta:J(\eta)<\infty\}$ or a subspace therein, and function
evaluations typically appear in $L(\eta|{\text{data}})$.  To
facilitate analysis and computation, one needs a metric and a geometry
in the function space, and needs the evaluation functional to be
continuous.

A reproducing kernel Hilbert space is a Hilbert space $\mathcal{H}$ of
functions on a domain $\mathcal{X}$ in which the evaluation functional
$[x]\eta=\eta(x)$ is continuous, $\forall{x}\in\mathcal{X}$,
$\forall\eta\in\mathcal{H}$.  By Riesz representation, there exists a
reproducing kernel, a non-negative definite bivariate function
$R(x,y)$ dual to the inner product $\langle\cdot,\cdot\rangle$ in
$\mathcal{H}$, which satisfies
$\langle{R}(x,\cdot),\eta(\cdot)\rangle=\eta(x)$,
$\forall{x}\in\mathcal{X}$, $\forall\eta\in\mathcal{H}$.  A
reproducing kernel Hilbert space can also be generated from its
reproducing kernel $R(x,y)$, for which {\it any} non-negative definite
function qualifies, as the ``column space''
$\text{span}\{R(x,\cdot),x\in\mathcal{X}\}$.

The space to use in (\ref{pelk}) is some
$\mathcal{H}\subseteq\big\{\eta:J(\eta)<\infty\big\}$ in which
$J(\eta)$ is a semi square norm, and there is a tensor-sum
decomposition $\mathcal{H}=\mathcal{N}_{J}\oplus\mathcal{H}_{J}$,
where the null space $\mathcal{N}_{J}=\{\eta:J(\eta)=0\}$ is of
finite-dimension, $\mathcal{H}_{J}$ has $J(\eta)$ as its full square
norm and the reproducing kernel $R_{J}(x,y)$ satisfying
$J\big(R_{J}(x,\cdot),\eta(\cdot)\big)=\eta(x)$,
$\forall\eta\in\mathcal{H}_{J}$; the square norm $\tilde{J}(\eta)$ of
$\mathcal{N}_{J}$ annihilates $\eta\in\mathcal{H}_{J}$.

For the cubic smoothing splines in (\ref{cubic}) and (\ref{logit}), a
configuration has
\begin{center}
  $\mathcal{H}_{J}=\{\eta:\int_{0}^{1}\ddot{\eta}^{2}(x)dx<0,
  \int_{0}^{1}\eta(x)dx=\int_{0}^{1}\dot{\eta}(x)dx=0\}$
\end{center}
with reproducing kernel $R_{J}(x,y)=k_{2}(x)k_{2}(y)-k_{4}(x-y)$,
where $k_{\nu}=B_{\nu}/\nu!$ are scaled Bernoulli polynomials.  The
null space can be further decomposed as
$\mathcal{N}_{J}=\{1\}\oplus\{k_{1}\}$, where $k_{1}(x)=x-0.5$, with
square norms $\tilde{J}_{0}(\eta)=\big(\int_{0}^{1}\eta(x)dx\big)^{2}$
and $\tilde{J}_{1}(\eta)=\big(\int_{0}^{1}\dot{\eta}(x)dx\big)^{2}$, in
order, and respective reproducing kernels $R_{00}(x,y)=1$ and
$R_{01}(x,y)=k_{1}(x)k_{1}(y)$.

A one-way ANOVA decomposition is built in, with the constant term in
$\{1\}$ and the contrast term in $\{k_{1}\}\oplus\mathcal{H}_{J}$
satisfying $\int_{0}^{1}\eta(x)dx=0$.

A comprehensive theory of reproducing kernel Hilbert spaces can be
found in \citeN{aron:50}.  Further details concerning the materials
presented here are in \citeN[Sect.~2.1,~2.3]{gu:13}

\subsection{Functional ANOVA and Tensor Product Splines}

On $\mathcal{X}=\mathcal{X}_{1}\times\mathcal{X}_{2}$, one has a
functional ANOVA decomposition of
$\eta(x)=\eta(x_{\sla1\sra},x_{\sla2\sra})$,
\begin{align}
\eta(x)&=(I-A_{1}+A_{1})(I-A_{2}+A_{2})\eta\notag\\
&=A_{1}A_{2}\eta+(I-A_{1})A_{2}\eta+A_{1}(I-A_{2})\eta+(I-A_{1})(I-A_{2})\eta\notag\\
\label{anova}
&=\eta_{\emptyset}+\eta_{1}(x_{\sla1\sra})+\eta_{2}(x_{\sla2\sra})
+\eta_{12}(x_{\sla1\sra},x_{\sla2\sra}),
\end{align}
where $I$ is the identity operator, $A_{1}$, $A_{2}$ are averaging
operators acting respectively on arguments $x_{\sla1\sra}$,
$x_{\sla2\sra}$ that satisfy $A1=1$.  Examples of averaging operators
include $Af=\int_{a}^{b}f(x)dx/(b-a)$ and
$Af=\sum_{i=1}^{m}f(x_{i})/m$.  Extensions to more than two dimensions
are straightforward.

Now consider $\mathcal{X}=[0,1]^{2}$.  Remember that reproducing
kernel Hilbert spaces can be generated from reproducing kernels.
Given non-negative definite functions
$R^{\sla1\sra}(x_{\sla1\sra},y_{\sla1\sra})$ on $\mathcal{X}_{1}$ and
$R^{\sla2\sra}(x_{\sla2\sra},y_{\sla2\sra})$ on $\mathcal{X}_{2}$,
$R(x,y)=R^{\sla1\sra}(x_{\sla1\sra},y_{\sla1\sra})
R^{\sla2\sra}(x_{\sla2\sra},y_{\sla2\sra})$ is non-negative definite
on $\mathcal{X}_{1}\times\mathcal{X}_{2}$.  Using marginal kernels
$R_{00}=1$, $R_{01}=k_{1}(x)k_{1}(y)$, and
$R_{1}=k_{2}(x)k_{2}(y)-k_{4}(x-y)$ on $[0,1]$, one may form nine
product kernels $R_{\alpha,\beta}$ on $[0,1]^{2}$ that generate nine
product spaces $\mathcal{H}_{\alpha,\beta}$:
\begin{center}
  \begin{tabular}{ccc}
    $R_{00,00}=R_{00}^{\sla1\sra}R_{00}^{\sla2\sra}$ &
    $R_{00,01}=R_{00}^{\sla1\sra}R_{01}^{\sla2\sra}$ &
    $R_{00,1}=R_{00}^{\sla1\sra}R_{1}^{\sla2\sra}$\\
    $R_{01,00}=R_{01}^{\sla1\sra}R_{00}^{\sla2\sra}$
    & $R_{01,01}=R_{01}^{\sla1\sra}R_{01}^{\sla2\sra}$ &
    $R_{01,1}=R_{01}^{\sla1\sra}R_{1}^{\sla2\sra}$\\
    $R_{1,00}=R_{1}^{\sla1\sra}R_{00}^{\sla2\sra}$
    & $R_{1,01}=R_{1}^{\sla1\sra}R_{01}^{\sla2\sra}$ &
    $R_{1,1}=R_{1}^{\sla1\sra}R_{1}^{\sla2\sra}$
  \end{tabular}
\end{center}
The four spaces on the upper left corner are of one dimension each, to
be lumped to form $\mathcal{N}_{J}$.  The tensor-sum of the other five
spaces can be used as $\mathcal{H}_{J}$ with reproducing kernel
\begin{center}
  $R_{J}=\theta_{00,1}R_{00,1}+\theta_{1,00}R_{1,00}+\theta_{01,1}R_{01,1}
  +\theta_{1,01}R_{1,01}+\theta_{1,1}R_{1,1}$,
\end{center}
where the $\theta$'s adjust the relative contributions of the
components to the overall roughness measure.  This yields tensor
product cubic splines.

The ANOVA decomposition of (\ref{anova}) is built in with
$\eta_{0}\in\mathcal{H}_{00,00}$,
$\eta_{1}\in\mathcal{H}_{01,00}\oplus\mathcal{H}_{1,00}$,
$\eta_{2}\in\mathcal{H}_{00,01}\oplus\mathcal{H}_{00,1}$, and
$\eta_{12}\in\mathcal{H}_{01,01}\oplus\mathcal{H}_{1,01}
\oplus\mathcal{H}_{01,1}\oplus\mathcal{H}_{1,1}$; this is inherited
from the one-way ANOVA decomposition on $[0,1]$.  To force an additive
model, one may remove $\mathcal{H}_{01,01}$ from $\mathcal{N}_{J}$ and
set $\theta_{1,01}=\theta_{01,1}=\theta_{1,1}=0$.

The above construction builds on reproducing kernels on marginal
domains, but the domain does not have to be $[0,1]^2$ and the marginal
configurations do not have to be cubic splines.  Some other
domain-kernel pairs can be found in \citeN[Sect.~2.2-2.4,
  Chap.~4]{gu:13}; see also \citeN{gw:22}.  The construction readily
extends to more than two dimensions.

With tensor product splines, one has
$\mathcal{H}_{J}=\oplus_{\beta}\mathcal{H}_{\beta}$ having square norm
$J(\eta)=\sum_{\beta}\theta_{\beta}^{-1}J_{\beta}(\eta_{\beta})$ and
reproducing kernel $R_{J}=\sum_{\beta}\theta_{\beta}R_{\beta}$, where
$\eta=\sum_{\beta}\eta_{\beta}$ for
$\eta_{\beta}\in\mathcal{H}_{\beta}$,
$J_{\beta}\big(R_{\beta}(x,\cdot),\eta_{\beta}(\cdot)\big)=\eta_{\beta}(x)$,
$\forall\eta_{\beta}\in\mathcal{H}_{\beta}$, and
$J_{\beta}(\eta_{\gamma})=0$, $\gamma\neq\beta$.  The
$\theta_{\beta}$'s are extra smoothing parameters to be selected along
with the $\lambda$ in front of $J(\eta)$; $(\lambda,\theta_{\beta})$ is an
overparameterization of $\lambda_{\beta}=\lambda/\theta_{\beta}$.

\subsection{Expressions of Estimates}

The space $\mathcal{H}$ is generally of infinite dimension, but when
$L(\eta|\text{data})$ depends on $\eta$ only through $\eta(x_{i})$,
$i=1,\dots,n$, the minimizer of (\ref{pelk}) has an expression
$\eta(x)=\sum_{\nu=1}^{m}d_{\nu}\phi_{\nu}(x)+\sum_{i=1}^{n}c_{i}R_{J}(x_{i},x)$,
where $\phi_{\nu}$'s span $\mathcal{N}_{J}$; see \citeN{kimel:71}.

It can be shown that the minimizer of (\ref{pelk}) in some
$\mathcal{H}^{*}=\mathcal{N}_{J}\oplus\text{span}\big\{R_{J}(z_{j},\cdot),j=1,\dots,q\big\}$
shares the same asymptotic convergence rates as that in $\mathcal{H}$,
where $q\asymp{n}^{\alpha}$ for some $\alpha\in(0,1)$ and $\{z_{j}\}$
is a random subset of $\{x_{i}\}$; see \citeN{gk:02}.

In practice, one calculates the minimizer in $\mathcal{H}^{*}$ with expression
\begin{equation}\label{expr}
  \textstyle
  \eta(x)=\sum_{\nu=1}^{m}d_{\nu}\phi_{\nu}(x)+\sum_{j=1}^{q}c_{j}R_{J}(z_{j},x).
\end{equation}
For $R_{J}=\sum_{\beta}\theta_{\beta}R_{\beta}$ as with tensor product
splines, $\eta=\eta_{0}+\sum_{\beta}\eta_{\beta}$, where
$\eta_{0}\in\mathcal{N}_{J}$,
$\eta_{\beta}=\theta_{\beta}\sum_{j=1}^{q}c_{j}R_{\beta}(z_{j},x)\in\mathcal{H}_{\beta}$,
and
$J_{\beta}(\eta_{\beta})=\theta_{\beta}^{2}\sum_{j,k}c_{j}c_{k}R_{\beta}(z_{j},z_{k})$.

\subsection{Kullback-Leibler Projection}\label{klproj}

ANOVA structures may be enforced via selective term elimination in
estimation, and may be inferred from fitted models containing possibly
redundant terms.  The latter task resembles hypothesis testing, with
$H_{0}:\eta\in\mathcal{H}_{0}$ versus
$H_{a}:\eta\in\mathcal{H}_{0}\oplus\mathcal{H}_{1}$, say; for an
example, consider
$\mathcal{H}_{0}=\{\eta:\eta=\eta_{\emptyset}+\eta_{1}+\eta_{2}\}$ and
$\mathcal{H}_{1}=\{\eta:\eta=\eta_{12}\}$ using the notation of
(\ref{anova}).

Lacking sampling distributions in settings with infinite-dimensional
nulls, the classical testing approach is of little help in this
situation.  Instead, an approach based on the Kullback-Leibler
geometry was developed in \citeN{gu:03}: one calculates an estimate
$\hat{\eta}\in\mathcal{H}_{0}\oplus\mathcal{H}_{1}$, obtains its
Kullback-Leibler projection $\tilde{\eta}\in\mathcal{H}_{0}$ by
minimizing a setting-specific ${\rm{KL}}(\hat{\eta},\eta)$ over
$\eta\in\mathcal{H}_{0}$, then inspects an ``entropy decomposition,''
%\begin{center}
${\rm{KL}}(\hat{\eta},\eta_{c})={\rm{KL}}(\hat{\eta},\tilde{\eta})
+{\rm{KL}}(\tilde{\eta},\eta_{c})$,
%\end{center}
an exact or approximate identity, where $\eta_{c}$ is a degenerate fit
such as a constant regression function.  When
$\gamma={\rm{KL}}(\hat{\eta},\tilde{\eta})/{\rm{KL}}(\hat{\eta},\eta_{c})$
is small, one loses little by cutting out $\mathcal{H}_{1}$.

\section{Smoothing Parameter Selection}

Varying the $\lambda$ in front of $J(\eta)$ and possible
$\theta_{\beta}$'s hidden therein, the minimizer of (\ref{pelk})
defines a family of point estimates given the data, from which one
needs to pick one member to use via smoothing parameter selection.

To estimate the unknown true $\eta$ using observed data, various
cross-validation schemes have been developed for smoothing parameter
selection; see, e.g., \citeN[Sect.~3.2,~5.2,~8.6]{gu:13}.  The aim is
to minimize appropriate statistical losses in respective settings,
such as the mean square error loss in Gaussian regression and the
Kullback-Leibler loss in non-Gaussian regression.

To define clones of the known master copy $\eta^{*}$ based on
simulated replicates, we shall pick minimizers of (\ref{pelk}) that
share certain characteristics with $\eta^{*}$.  Statistical estimates
are compromises between the data and the model constraints, and our
choice of the matching characteristics are the effective constraints.

\subsection{Effective Constraints and Optimal Performer}\label{eta1eta2}

The penalized likelihood of (\ref{pelk}) is effectively performing
constrained maximum likelihood estimation,
\begin{center}
  $\min\,L(\eta|\text{data})\qquad\quad{s.t.}\quad J(\eta)=\rho$,
\end{center}
using the Lagrange method; see \citeN[Thm.~2.12]{gu:13}.

For $J(\eta)$ without $\theta_{\beta}$'s hidden in, such as with the
cubic splines of (\ref{cubic}) and (\ref{logit}), the constraint
$J(\eta)=\rho$ acts just like a parametric model.  Given
$J(\eta^{*})=\rho^{*}$, the clones should satisfy $J(\eta)=\rho^{*}$.
This is the familiar notion of estimator, a collection of estimates
subject to the same set of model constraints based on samples from the
same source.

While the mapping $\lambda\leftrightarrow\rho$ is one-to-one given the
data, it does vary with $L(\eta|\text{data})$.  The same $\rho$
corresponds to different $\lambda$'s for different replicates
simulated from the same source.  Fixed $\lambda$ in (\ref{pelk}) does
not define an estimator, and clones should not be produced by any
fixed $\lambda$.

\begin{figure}
\centerline{\includegraphics[width=.2\linewidth,height=\linewidth,angle=270]{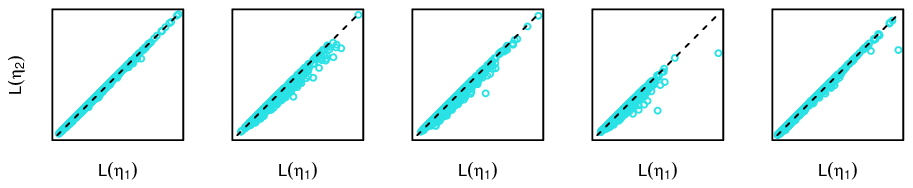}}
\caption{Performances of Effective Constraints versus Optimal
  Performer.  From left to right, (i) Gaussian regression simulation,
  (ii) logistic regression simulation, (iii) additive fit to {\tt
    LakeAcidity} from Section~4.2, (iv) additive fit to {\tt wesdr}
  from Section~4.3, and (v) Weibull fit to {\tt stan} from
  Section~4.4.  The frames contain one thousand replicates each.}
\label{fig1}
\end{figure}

Some simple simulations were conducted with $n=100$,
$x_{i}\sim{U}(0,1)$, $\eta(x)=1+3\sin(2\pi{x}-\pi)$, and
$\epsilon_{i}\sim{N}(0,1)$ in (\ref{cubic}).  One set of
$(x_{i},Y_{i})$ pairs were generated as observed data, and a point
estimate $\eta^{*}$ was calculated with $q=30$ $z_{j}$'s in
(\ref{expr}) and with $\lambda$ selected by generalized
cross-validation of \citeN{craven:79}; a variance estimate
$\tilde{\sigma}^{2}$ was also obtained.  One thousand replicates were
then simulated via
$\tilde{Y}_{i}\sim{N}\big(\eta^{*}(x_{i}),\tilde{\sigma}^{2})$, and
for each replicate of $n=100$ $(x_{i},\tilde{Y}_{i})$ pairs, two
minimizers of (\ref{cubic}) were calculated via the selection of
$\lambda$, with $\eta_{1}$ satisfying $J(\eta_{1})=\rho^{*}$ and
$\eta_{2}$ attaining the minimum mean square error loss
$L(\eta)=\frac{1}{n}\sum_{i=1}^{n}\big(\eta(x_{i})-\eta^{*}(x_{i})\big)^{2}$;
the same set of $z_{j}$'s were used as in $\eta^{*}$.  $L(\eta_{1})$
versus $L(\eta_{2})$ for the thousand replicates are plotted in the
first frame of Figure~\ref{fig1}.

Parallel simulations were also conducted with
$Y_{i}\sim\text{Bin}\big(3,p(x_{i})\big)$ in (\ref{logit}) for
$p=\frac{e^{\eta}}{1+e^{\eta}}$ and
$\eta(x)=3\{10^{5}x^{11}(1-x)^{6}+10^{3}x^{3}(1-x)^{10}\}-2$; direct
cross-validation in \citeN[Sect.~5.4.2]{gu:13} was used to obtain
$\eta^{*}$.  One thousand replicates were simulated with
$\tilde{Y}_{i}\sim\text{Bin}\big(3,p^{*}(x_{i})\big)$, and for each
replicate, minimizer $\eta_{1}$ of (\ref{logit}) satisfies
$J(\eta_{1})=\rho^{*}$, and minimizer $\eta_{2}$ attains the minimum
Kullpack-Leibler loss
\begin{center}
  $L(\eta)=\frac{1}{n}\sum_{i=1}^{n}m_{i}\big\{p^{*}(\eta^{*}-\eta)
  -\log(1+e^{\eta^{*}})+\log(1+e^{\eta})\big\}(x_{i})$,
\end{center}
where $m_{i}=3$.  The respective $L(\eta_{1})$ versus $L(\eta_{2})$
comparison over these replicates are shown in the second frame of
Figure~\ref{fig1}.

As noted above, $\eta_{1}$ should be taken as the clone to use.  The
proposed paradigm is open to all estimation methods, but in settings
where effective constraints are not explicitly available or
quantifiable, one may need a fallback option; given the observed near
lockstep between $L(\eta_{1})$ and $L(\eta_{2})$, the optimal
performer $\eta_{2}$, always available in principle, might serve as
clones.  We shall refer to $\eta_{1}$ and $\eta_{2}$ as clones of type
I and type II, respectively.

Note that $\eta_{1}$ and $\eta_{2}$ may not be ``physically'' close to
each other even if $L(\eta_{1})\approx{L}(\eta_{2})$, as the bottom of
$L(\eta)$ could be flat.

Some closely related conceptual discussions can be found in
\citeN{gu:98b}, concerning proper model indexing (i.e., the definition
of estimator) in nonparametric settings.

\subsection{Multiple Smoothing Parameters}

For $J(\eta)=\sum_{\beta}\theta_{\beta}^{-1}J_{\beta}(\eta_{\beta})$
with tensor product splines, $\lambda_{\theta}=\lambda/\theta_{\beta}$
are selected via cross-validation to obtain
$\eta^{*}=\eta_{0}^{*}+\sum_{\beta}\eta_{\beta}^{*}$, where
$\eta_{0}^{*}\in\mathcal{N}_{J}$,
$\eta_{\beta}^{*}\in\mathcal{H}_{\beta}$.

Ideally, clones of form $\eta=\eta_{0}+\sum_{\beta}\eta_{\beta}$
should satisfy
$J_{\beta}(\eta_{\beta})=\rho_{\beta}^{*}=J_{\beta}(\eta_{\beta}^{*})$,
the effective constraints $\eta^{*}$ satisfies, which however may not
be feasible to execute.  An operable proxy is to select
$\lambda_{\beta}$'s for $\eta_{1}$ via the minimization of
$\sum_{\beta}w_{\beta}\big(\log{J}_{\beta}(\eta_{\beta})-\log\rho_{\beta}^{*}\big)^{2}$,
with weights $w_{\beta}$ proportional to the sample variances of
$\eta_{\beta}^{*}(x_{i})$, say; other weights were also tried, such as
the square root of the current choice, with negligible discrepancies
seen in the resulting $L(\eta_{1})$.

Statistical losses are not affected by the number of smoothing
parameters, so the definition of optimal performer $\eta_{2}$ remains
the same.

\section{Examples}

We now demonstrate retrospective inference through a few simulated and
real-data examples.

{\tt R} package {\tt gss} implements a comprehensive set of tools for
smoothing spline ANOVA models in the settings of Gaussian and
non-Gaussian regression, density estimation, and hazard estimation;
see \citeN{gss:14}.  We shall use the previously existing {\tt gss}
facilities in the regression settings, plus some newly added tools for
the proposed retrospective inference.

\subsection{Cubic Spline Simulations}

First look at the cubic spline simulations of Section~\ref{eta1eta2}.
\begin{figure}[t]
\centerline{\includegraphics[height=.9\linewidth,width=.6\linewidth,angle=270]{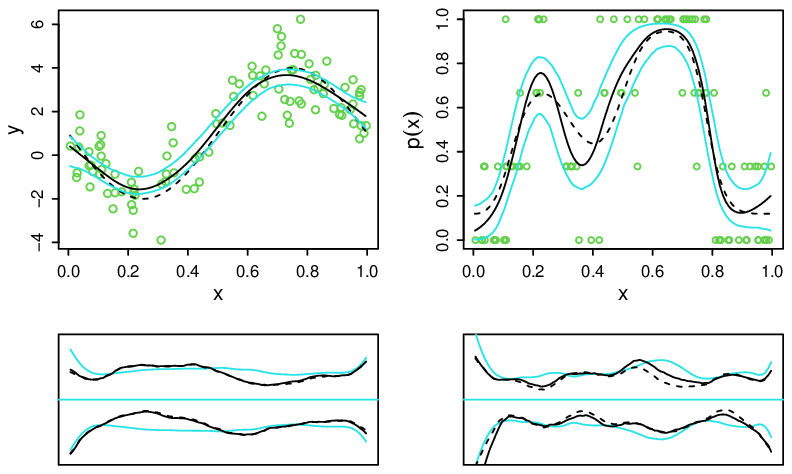}}
\caption{Cubic Spline Simulations.  Top: Cross-validated $\eta^{*}$
  are in solid lines, true $\eta$ in dashed lines, 95\% retrospactive
  confidence band (using type I clones) in faded lines, and observed
  data in circles.  Bottom: Confidence bands from clones of type I
  (solid lines) and type II (dashed lines), on $\eta$ scale; Bayesian
  confidence intervals are in faded lines and horizontal lines
  represent $\eta^{*}$.  Gaussian regression is on the left and
  logistic regression on the right.}
\label{fig2}
\end{figure}

Plotted in the top-left frame of Figure~\ref{fig2} is a
cross-validated point estimate $\eta^{*}$ (solid line) using observed
data (circles) in the Gaussian regression of (\ref{cubic}); the true
$\eta$ is superimposed in the dashed line.  Clones of type I and type
II were calculated using 1,000 replicates simulated via
$\tilde{Y}_{i}\sim{N}\big(\eta^{*}(x_{i}),\tilde{\sigma}^{2}\big)$.
Evaluating these clones at any $x\in[0,1]$, one can obtain quantiles
of the 1,000 resulting function values, and a 95\% retrospective
confidence band, based on the type I clones, is plotted in faded lines
by connecting the $(2.5\%,97.5\%)$ quantiles on a grid; replacing
$(2.5\%,95.5\%)$ by $(0\%,100\%)$, one may obtain an envelop
encapsulating all 1,000 clones, though the theoretical limits of a
100\% confidence band should be $(-\infty,\infty)$.

Note that no multiplicity adjustments are needed here going from
confidence intervals to confidence bands, as all are just different
``slices'' of the same clone distribution.

The bottom-left frame of Figure~\ref{fig2} compares the 95\%
confidence bands based on clones of type I (solid lines) and type II
(dashed lines) after subtracting $\eta^{*}$; the 95\% Bayesian
confidence intervals of \citeN{wahba:83} are also superimposed in
faded lines.

Parallel results in the logistic regression of (\ref{logit}) are shown
in the right frames of Figure~\ref{fig2}, with the top frame on the
probability scale and the bottom frame on the logit scale.

By construction, Bayesian confidence intervals are symmetric on the
upper and lower sides of $\eta^{*}$.  The retrospective confidence
bands are asymmetric in general, seeming to lean downwards at peaks of
$\eta^{*}$ and upwards at valleys; the medians of the clone
evaluations are likely flatter than $\eta^{*}$.

\subsection{Water Acidity in Lakes}

Some data extracted from the Eastern Lake Survey of 1984 conducted by
the United States Environmental Protection Agency are included in {\tt
  gss}, concerning 112 lakes in the Blue Ridge.  Detailed analysis of
the data can be found in \citeN[Sect.~4.3.4]{gu:13} using tools
available then.

The {\tt R} code below load package {\tt gss} and data frame {\tt
  LakeAcidity}, then fit a smoothing spline ANOVA model to the data:
\begin{quote}
\begin{verbatim}
library(gss); data(LakeAcidity)
fit.lake <- ssanova(ph~log(cal)*geog,data=LakeAcidity,seed=5732)
\end{verbatim}
\end{quote}
This is Gassian regression with surface pH as the response, calcium
concentration and geographic location as covariates, and {\tt seed}
ensures reproducible selection of $z_{j}$'s in (\ref{expr}).

The fit is a tensor product spline, with cubic spline marginal on {\tt
  log(cal)} and thin-plate spline marginal on {\tt geog}, where {\tt
  geog} is the $x$-$y$ coordinate (in distance) with respect to a
local origin converted from longitude-latitude.  Thin-plate splines on
$R^{2}$ are invariant to coordinate shift and rotation, and the
mathematically two-dimensional {\tt geog} contributes one logical
dimension in tensor product spline; see \citeN[Sect.~4.3]{gu:13}.

The ANOVA decomposition of (\ref{anova}) is built in, with terms
labeled {\tt "1"}, {\tt "log(cal)"}, {\tt "geog"}, and {\tt
  "log(cal):geog"}.  To assess the adequacy of an additive model, one
may calculate the Kullback-Leibler projection of Section~\ref{klproj}
to obtain the ratio
$\gamma=\text{KL}(\hat{\eta},\tilde{\eta})/\text{KL}(\hat{\eta},\eta_{c})
=0.00056$ along with a retrospective $p$-value $0.925$:
\begin{quote}
\begin{verbatim}
project1(fit.lake,include=c("log(cal)","geog")) 
\end{verbatim}
\end{quote}

Remember that the point estimate
$\eta^{*}=\hat{\eta}\in\mathcal{H}_{0}\oplus\mathcal{H}_{1}$ in the
notation of Section~\ref{klproj} and $\tilde{\eta}\in\mathcal{H}_{0}$
is the projection of $\eta^{*}$ in $\mathcal{H}_{0}$.  For the purpose
of calibrating the $\gamma$ ratio, we use $\tilde{\eta}$ as the truth
to simulate replicates, calculate clones in
$\mathcal{H}_{0}\oplus\mathcal{H}_{1}$, then obtain the $\gamma$
ratios for the clones; such clones will be called clones under the
null.  By default, {\tt project1} simulates 200 clones under the null,
with the retrospective $p$-value being the percent of cloned ratios
larger than the observed $\gamma$.

An additive model can now be fitted to the data with $q=n$ in
(\ref{expr}), along with 1,000 clones:
\begin{quote}
\begin{verbatim}
fit.lake.a <- ssanova(ph~log(cal)+geog,data=LakeAcidity,id.basis=1:112)
cln.lake <- clone(fit.lake.a)
\end{verbatim}
\end{quote}
These clones are based on replicates simulated with the additive
$\eta^{*}$ as the truth, and the default is of type I.  To obtain
clones of type II, simply use {\tt clone(fit,type=2)}.

Terms {\tt "log(cal)"} and {\tt "geog"} can be evaluated separately
and could have their respective retrospective confidence bands.  One
may evaluate the {\tt "log(cal)"} term on the data points and the {\tt
  "geog"} term on a grid:
\begin{quote}
\begin{verbatim}
est.c <- predict(fit.lake.a,fit.lake.a$mf,inc="log(cal)")
rci.c <- retroCI(cln.lake,fit.lake.a$mf,inc="log(cal)")
grid0 <- seq(-.04,.04,len=31)
grid <- cbind(rep(grid0,31),rep(grid0,rep(31,31)))
est.g <- predict(fit.lake.a,data.frame(geog=I(grid)),inc="geog")
rci.g <- retroCI(cln.lake,data.frame(geog=I(grid)),inc="geog")
\end{verbatim}
\end{quote}
where {\tt retroCI} returns a matrix of two columns of the
$(2.5\%,97.5\%)$ quantiles of clone evaluations.

\begin{figure}[t]
\centerline{\includegraphics[height=.8\linewidth,width=.8\linewidth,angle=270]{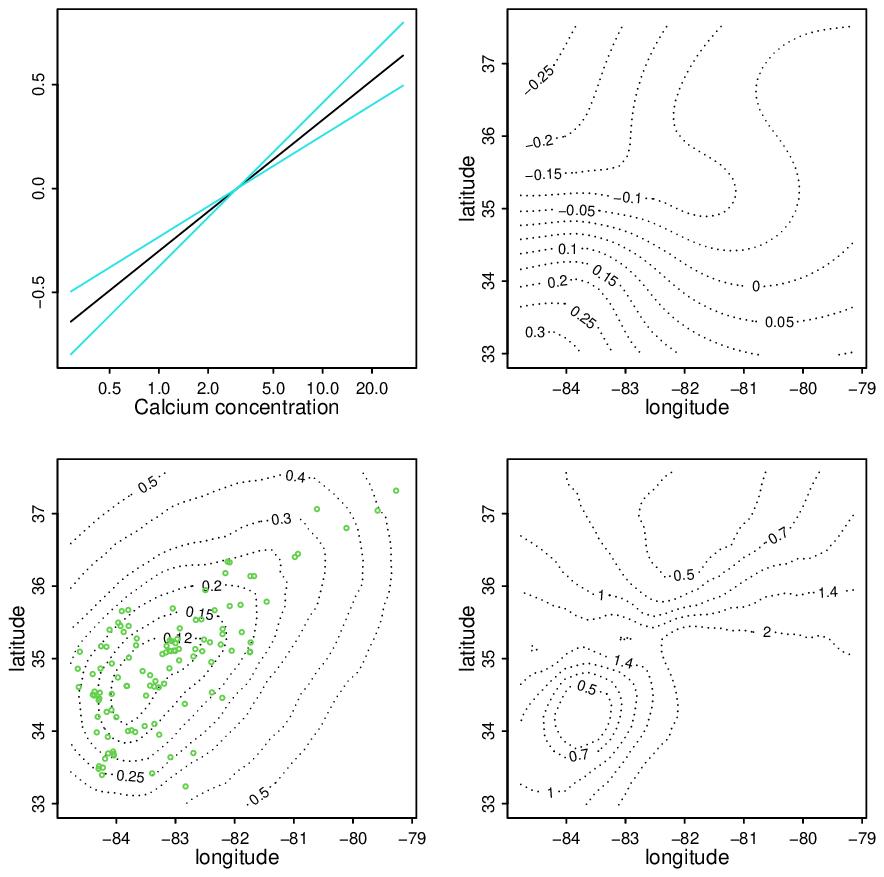}}
\caption{Additive Fit to LakeAcidity.  Top-left: Estimated {\tt
    "log(cal)"} with 95\% retrospective confidence band.  Top-right:
  Estimated {\tt "geog"}.  Bottom-left: Width of 95\% {\tt "geog"}
  retrospective confidence band with lakes superimposed.
  Bottom-right: Asymmetry of 95\% {\tt "geog"} retrospective
  confidence band.}
\label{fig3}
\end{figure}
Shown in Figure~\ref{fig3} are (i) the estimated {\tt "log(cal)"} term
with 95\% retrospective confidence band (top left), (ii) the estimated
{\tt "geog"} term (top right), (iii) the width of 95\% retrospective
confidence band of {\tt "geog"} (bottom left), {\tt
  rci.g[,2]-rci.g[,1]}, and (iv) the asymmetry of 95\% retrospective
confidence band of {\tt "geog"} (bottom right), {\tt
  (rci.g[,2]-est.g)/(est.g-rci.g[,1])}; the grid used in {\tt "geog"}
evaluations is converted back to longitude-latitude in the plots.

$L(\eta_{1})$ versus $L(\eta_{2})$ for a thousand replicates from the
additive fit are shown in the third frame of Figure~\ref{fig1}.

\subsection{Progression of Diabetic Retinopathy}

Also included in {\tt gss} are some data collected on 669 patients
from the Wisconsin Epidemiological Study of Diabetic Retinopathy.
Some analysis of the data can be found in \citeN[Sect.~5.5.3]{gu:13}.

The {\tt R} code below load package {\tt gss} and data frame {\tt
  wesdr}, then fit a smoothing spline ANOVA logistic model to the
data:
\begin{quote}
\begin{verbatim}
library(gss); data(wesdr)
fit.wsd <- gssanova(ret~dur*bmi*gly,"binomial",data=wesdr,seed=5732)
\end{verbatim}
\end{quote}
The response is a binary indicator of retinopathy progression at the
first follow-up and the covariates are baseline measures of duration
of diabetes in years, body mass index, and percent of glycosylated
hemoglobin.

\begin{figure}[t]
\centerline{\includegraphics[height=\linewidth,width=.5\linewidth,angle=270]{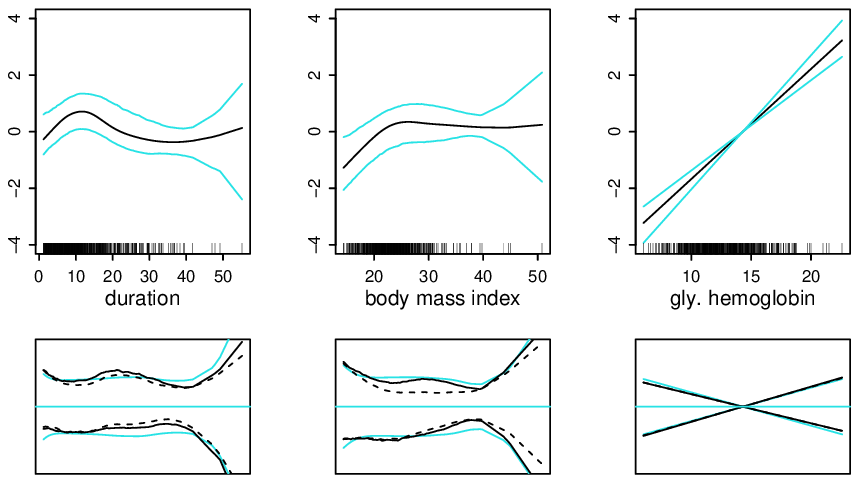}}
\caption{Additive Fit to {\tt wesdr}.  Top: Terms {\tt "dur"}, {\tt
    "bmi"}, and {\tt "gly"} in cross-validated $\eta^{*}$ with 95\%
  retrospective confidence bands; rugs on bottom mark data points.
  Bottom: Confidence bands from clones of type I (solid lines) and
  type II (dashed lines); Bayesian confidence intervals are in faded
  lines and horizontal lines represent terms in $\eta^{*}$.}
\label{fig4}
\end{figure}

The fit is a tensor product cubic spline, containing three main
effects, three two-way interactions, and a three-way interaction, for
a total of 19 $\theta_{\beta}$'s in $J(\eta)$.  The Kullback-Leibler
projection to an additive model has a $\gamma=0.027$ with
retrospective $p$-value $0.345$:
\begin{quote}
\begin{verbatim}
project1(fit.wsd,c("dur","bmi","gly"))
\end{verbatim}
\end{quote}
An additive model can then be fitted and 1,000 clones produced:
\begin{quote}
\begin{verbatim}
fit.wsd.a <- gssanova(ret~dur+bmi+gly,"binomial",data=wesdr,seed=5732)
cln.wsd <- clone(fit.wsd.a)
\end{verbatim}
\end{quote}
Terms {\tt "dur"}, {\tt "bmi"}, and {\tt "gly"} in the additive
$\eta^{*}$ can be evaluated at the data points and respective 95\%
retrospective confidence bands obtained:
\begin{quote}
\begin{verbatim}
est.dur <- predict(fit.wsd.a,wesdr,inc="dur")
rci.dur <- retroCI(cln.wsd,wesdr,inc="dur")
est.bmi <- predict(fit.wsd.a,wesdr,inc="bmi")
rci.bmi <- retroCI(cln.wsd,wesdr,inc="bmi")
est.gly <- predict(fit.wsd.a,wesdr,inc="gly")
rci.gly <- retroCI(cln.wsd,wesdr,inc="gly")
\end{verbatim}
\end{quote}
The results are shown in Figure~\ref{fig4} in the same manner as
Figure~\ref{fig2}, but with everything on the logit scale; ANOVA terms
only make sense on the link scale.

$L(\eta_{1})$ is plotted against $L(\eta_{2})$ in the fourth frame of
Figure~\ref{fig1} over a thousand replicates from the additive fit.

\subsection{Survival After Heart Transplant}

Let $T$ be the lifetime of an item with survival function
$S(t|u)=P(T>t|u)$, with $u$ a covariate, of interest is the estimation
of the hazard function $e^{\eta(t,u)}=-\log{S}(t|u)/dt$.  One may
write $\eta=\eta_{\emptyset}+\eta_{t}+\eta_{u}+\eta_{t,u}$ as in
(\ref{anova}); setting $\eta_{t,u}=0$, one has a proportional hazard
model with base hazard $e^{\eta_{\emptyset}+\eta_{t}}$ and relative
risk $e^{\eta_{u}}$.

One of the most demonstrated survival data is the Stanford heart
transplant data.  The data were used in a few case studies in
\citeN{gu:13}, (i) with $\eta(t,u)$ fully nonparametric (Sect.~8.4.2),
(ii) with $\eta(t,u)$ parametric in $t$ (Sect.~8.6.6), and (iii) the
estimation of relative risk $e^{\eta_{u}}$ via penalized partial
likelihood.

The {\tt R} code below load package {\tt gss} and data frame {\tt
  stan}, then fit a Weibull hazard to the data:
\begin{quote}
\begin{verbatim}
library(gss); data(stan)
fit.stan<-gssanova(cbind(time+.01,status)~age,"weibull",
                   data=stan,nbasis=200)
\end{verbatim}
\end{quote}
The follow-up times were rounded to whole days and there was a 0, and
we choose to add 0.01 to all instead of deleting the 0; {\tt status}
is the censoring indicator with 113 observed deaths and 71 censorings,
age at transplant is the covariate $u$, and {\tt nbasis=200} sets
$q=n$ in (\ref{expr}).  This is case study (ii) mentioned above.

\begin{figure}[t]
\centerline{\includegraphics[height=.45\linewidth,width=.6\linewidth,angle=270]{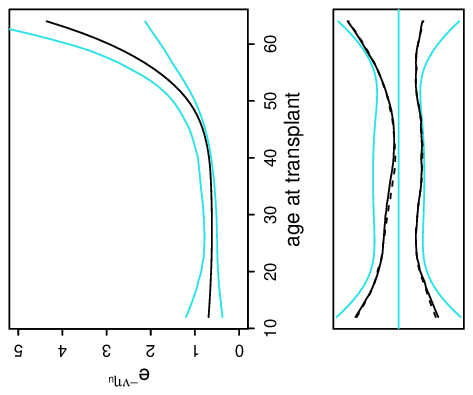}}
\caption{Weibull Fit to {\tt stan}.  Top: Relative risk using
  cross-validated $\eta^{*}_{u}$ with 95\% retrospective confidence
  band.  Bottom: Confidence bands of $\eta^{*}_{u}$ from clones of
  type I (solid lines) and type II (dashed lines); Bayesian confidence
  intervals are in faded lines and horizontal line represents
  $\eta^{*}_{u}$.}
\label{fig5}
\end{figure}

The Weibull hazard is of form
$e^{\eta(t,u)}=\frac{\nu}{t}\exp\{\nu(\log{t}-\eta(u))\}$, so it is a
proportional hazard model with $\exp\{-\nu\eta(u)\}$ proportional to
the relative risk.  The Kullback-Leibler loss $L(\eta)$ in the setting
and the associated cross-validation scheme can be found in
\citeN[Sect.~8.6.3]{gu:13}.  The cubic spline fit has a one-way ANOVA
decomposition built in, $\eta=\eta_{\emptyset}+\eta_{u}$, as noted in
Section~\ref{rkhs}.  The cross-validated $\eta^{*}_{u}$ and the
retrospective confidence band can be obtained on a grid:
\begin{quote}
\begin{verbatim}
grid<-seq(min(stan$age),max(stan$age),length=51)
est.age<-predict(fit.stan,data.frame(age=grid),inc="age")
cln.stan<-clone(fit.stan)
rci.age<-retroCI(cln.stan,data.frame(age=grid),inc="age")
\end{verbatim}
\end{quote}
The results are shown in Figure~\ref{fig5} in the same manner as
Figure~\ref{fig2}.

The fifth frame of Figure~\ref{fig1} compares $L(\eta_{1})$ and
$L(\eta_{2})$ in a thousand replicates from the fit.

\section{Parametric Regression}

We now put retrospective inference to work in parametric regression
and compare with traditional inference.  There are no tuning
parameters to worry about here, and the mapping from replicates to
clones has no ambiguity.

\subsection{Gaussian Regression}

Consider linear model $\sbf{Y}=X\sbf{\beta}+\sbf{\epsilon}$,
$\sbf{\epsilon}\sim{N}(\sbf{0},\sigma^{2}I)$.  Given the least squares
estimate $\sbf{\beta}^{*}$ and the variance estimate
$\tilde{\sigma}^{2}$, the replicates are to be generated via
$\tilde{\sbf{Y}}\sim{N}(X\sbf{\beta}^{*},\tilde{\sigma}^{2}I)$, and
the clone distribution is known to be
$\sbf{\beta}\sim{N}\big(\sbf{\beta}^{*},\tilde{\sigma}^{2}(X^{T}X)^{-1}\big)$.
Retrospective confidence intervals of $\sbf{x}^{T}\sbf{\beta}$ are
thus $\sbf{x}^{T}\sbf{\beta}^{*}
\pm{z}_{1-\alpha/2}\tilde{\sigma}\sqrt{\sbf{x}^{T}(X^{T}X)^{-1}\sbf{x}}$,
$\forall\sbf{x}$.  Apart from the use of the $z$-table instead of the
$t$-table, the construction is identical to that of traditional
confidence intervals, but no multiplicity adjustments are needed going
from a confidence interval to a confidence band.

Writing $X\sbf{\beta}=X_{1}\sbf{\beta}_{1}+X_{2}\sbf{\beta}_{2}$, we
shall now test the hypothesis $\sbf{\beta}_{2}=\sbf{0}$.  The entropy
decomposition
$\text{KL}(\hat{\eta},\eta_{c})=\text{KL}(\hat{\eta},\tilde{\eta})
+\text{KL}(\tilde{\eta}.\eta_{c})$ in Section~\ref{klproj} becomes
\[
\sum_{i}(\hat{Y}_{i}-\bar{Y})^{2}=\sum_{i}(\hat{Y}_{i}-\tilde{Y}_{i})^{2}
+\sum_{i}(\tilde{Y}_{i}-\bar{Y})^{2},
\]
where $\hat{\sbf{Y}}=X(X^{T}X)^{-1}X^{T}\sbf{Y}=P_{X}\sbf{Y}$,
$\tilde{\sbf{Y}}=X_{1}(X_{1}^{T}X_{1})^{-1}X_{1}^{T}\sbf{Y}=P_{X_{1}}\sbf{Y}$,
and
\[
\text{KL}(\hat{\eta},\tilde{\eta})=\sum_{i}(\hat{Y}_{i}-\tilde{Y}_{i})^{2}
=\sbf{Y}^{T}(P_{X}-P_{X_{1}})\sbf{Y}.
\]
Simulating replicates via
$\breve{\sbf{Y}}=\tilde{\sbf{Y}}+\tilde{\sbf{\epsilon}}$,
$\tilde{\sbf{\epsilon}}\sim{N}(\sbf{0},\tilde{\sigma}^{2}I)$,
$\breve{\sbf{Y}}^{T}(P_{X}-P_{X_{1}})\breve{\sbf{Y}}
=\tilde{\sbf{\epsilon}}^{T}(P_{X}-P_{X_{1}})\tilde{\sbf{\epsilon}}
\sim\tilde{\sigma}^{2}\chi^{2}_{\nu}$, where $\nu$ is the dimension of
$\sbf{\beta}_{2}$.  Using $\text{KL}(\hat{\eta},\tilde{\eta})$ as the
test statistic, its calibration via clone distribution under the null
goes through the $\chi^{2}$-table; note that
$\sum_{i}(\hat{Y}_{i}-\tilde{Y}_{i})^{2}=\text{SSR}(X_{2}|X_{1})$ and
$\tilde{\sigma}^{2}=\text{MSE}$ in standard linear model notation, so
one essentially works with the traditional test statistic
$F=\text{MSR}(X_{2}|X_{1})/\text{MSE}$, but calibrates it against
$F_{\nu,\infty}$.

The ratio
$\text{KL}(\hat{\eta},\tilde{\eta})/\text{KL}(\hat{\eta},\eta_{c})
=\text{KL}(\hat{\eta},\tilde{\eta})/\big(\text{KL}(\hat{\eta},\tilde{\eta})
+\text{KL}(\tilde{\eta},\eta_{c})\big)$ we choose to work with in {\tt
  project1} (see Section~4.2) has an intuitive information geometric
meaning, and it is nearly a monotone transform of
$\text{KL}(\hat{\eta},\tilde{\eta})$, except that
$\text{KL}(\tilde{\eta},\eta_{c})$ varies over clones under the null.
The ratio can be calibrated directly without reference to
probabilities, whereas $p$-values generally decrease as the observed
sample sizes increase.

\subsection{Non-Gaussian Regression}

In non-Gaussian regression, traditional inference largely relies on
asymptotic approximations, such as the normal approximation of
parameter estimates and the $\chi^{2}$ approximation of log likelihood
ratio.

Given a minus log likelihood $l(\sbf{\theta}|\text{data})$, its
minimizer $\hat{\sbf{\theta}}$, the maximum likelihood estimate, is
usually asymptotically normal,
$\hat{\sbf{\theta}}\stackrel{asy.}{\sim}{N}\big(\sbf{\theta}_{0},I^{-1}(\sbf{\theta}_{0})\big)$,
where $\sbf{\theta}_{0}$ is the truth and
$I(\sbf{\theta})=\partial^{2}l/\partial\sbf{\theta}\partial\sbf{\theta}^{T}$
is the information matrix.  To construct a 95\% confidence interval
for $\sbf{c}^{T}\sbf{\theta}$, say, one may use
$\sbf{c}^{T}\hat{\sbf{\theta}}\pm1.96\sqrt{\sbf{c}^{T}I^{-1}(\hat{\sbf{\theta}})\sbf{c}}$,
where the unknown $I(\sbf{\theta}_{0})$ is approximated by
$I(\hat{\sbf{\theta}})$ assuming a consistent $\hat{\sbf{\theta}}$.
If one chooses to use the asymptotic normal approximation for the
clone distribution with $\sbf{\theta}^{*}=\hat{\sbf{\theta}}$,
retrospective confidence intervals would be identical to traditional
confidence intervals.

For the testing of $\sbf{\beta}_{2}=\sbf{0}$ in a generalized linear
model $\eta(x)=\sbf{x}^{T}\sbf{\beta}=\sbf{x}_{1}^{T}\sbf{\beta}_{1}
+\sbf{x}_{2}^{T}\sbf{\beta}_{2}$, and with $\eta$ the canonical
parameter of an exponential family distribution,
\begin{center}
  $\text{KL}(\hat{\eta},\eta)=\sum_{i}\big\{-\hat{\mu}_{i}(\eta_{i}-\hat{\eta}_{i})
  +\big(b(\eta_{i})-b(\hat{\eta}_{i})\big)\big\}=l(\eta)-l(\hat{\eta})$,
\end{center}
where $g_{i}=g(x_{i})$ for any function $g$,
$\mu_{i}=\dot{b}(\eta_{i})$, and
$l(\eta)=\sum_{i}\{-y_{i}\eta_{i}+b(\eta_{i})\}$ is the minus log
likelihood; to see the last equation, differentiate
$l(\hat{\eta}+\alpha{h})$ with respect to scalar $\alpha$ and set the
derivative at $\alpha=0$ to zero, for all $h(x)$ in the model space,
noting that $\hat{\eta}$ minimizes $l(\eta)$.  Hence, $\tilde{\eta}$
is simply the reduced model fit and
$2\text{KL}(\hat{\eta},\tilde{\eta})$ the deviance difference between
$\hat{\eta}$ and $\tilde{\eta}$.  If one resorts to the asymptotic
$\chi^{2}$ approximation in such a setting, then the retrospective
$p$-value of $\text{KL}(\hat{\eta},\tilde{\eta})$ would be the same as
the $p$-value of the traditional likelihood ratio test.

For an illustration, consider again the {\tt wesdr} data used in
Section~4.3, but we now fit linear logistic models using {\tt glm}:
\begin{quote}
\begin{verbatim}
fit0<-glm(ret~dur*bmi*gly,"binomial",wesdr)  
fit1<-glm(ret~dur+bmi+gly,"binomial",wesdr)
\end{verbatim}
\end{quote}
Ten thousands replicates were generated via
$\tilde{y}_{i}\sim\text{Bin}(1,p_{1i})$ to produce clones of the
additive fit, for $p_{1i}=p_{1}(x_{i})$ obtained from {\tt fit1};
clones are computationally more affordable in parametric settings.  In
Table~\ref{tbl1}, the 95\% confidence intervals of regression
coefficients based on the clones are compared with those based on
asymptotic normality.  The clone-based and asymptotics-based 95\%
confidence intervals for the logit at the first data point {\tt
  wesdr[1,]} are respectively $(-0.0635,0.3310)$,
$(-0.0648,0.3217)$, translating to $(0.4841,0.5820)$,
$(0.4838,0.5798)$ on the probability scale.
\begin{table}\centering
  \begin{tabular}{r|cccc}
    & intercept & dur & bmi & gly\\\hline
    clone & $(-8.379,-5.306)$ & $(-0.0284,0.0121)$ & $(0.0267,0.1119)$ & $(0.3170,0.4758)$\\
    normal & $(-8.238,-5.204)$ & $(-0.0277,0.0124)$ & $(0.0252,0.1093)$ & $(0.3107,0.4672)$
  \end{tabular}
  \caption{Linear Additive Fit to {\tt wesdr}.  Comparisons of 95\%
    confidence intervals of regression coefficients, simulated clones
    versus asymptotic normality.}
  \label{tbl1}
\end{table}

For the testing of an additive model null, the above clones were
clones under the null; $2\text{KL}(\hat{\eta},\tilde{\eta})$ of the
clones followed $\chi^{2}_{4}$ reasonably well.  The observed
$2\text{KL}(\eta_{0},\eta_{1})=5.228$ has $p$-values $0.265$ from the
$\chi^{2}$-table and $0.278$ from the clone distribution under the
null.

The sample size 669 is large and the asymptotics seems to work well
here.

\section{Discussions}

In this article, we explore a new paradigm for statistical inference.
The central consideration is to nullify the necessity of prospective
finite-dimensional model constraints that are not available in
nonparametric settings.  The work flow is simple, intuitive, and
executable in parametric and nonparametric settings alike.

With no probabilistic claims concerning the truth, the exercise is
essentially a sensitivity analysis, but this may not be a bad thing.
To justify the probabilistic claims in traditional inference, one has
to live in a hypothetical world, arguing that the coverage probability
would materialize were the procedure applied over and over in similar
situations, or the like.  In the real world, however, one typically
has just the one data set at hand, and needs some uncertainty
quantification for the empirical findings drawn from the data.

As noted in Section~5, retrospective inference often returns the same
or similar numerical results as traditional inference (when the latter
is feasible); the difference is in the interpretation.  With
retrospective inference, asymptotic approximations could be avoided if
accuracy is in doubt, no multiplicity adjustments are needed going
from a confidence interval to a confidence band, and $p$-values are
actual probabilities.

The clone definition via effective constraints derives from the
familiar notion of estimator.  The notion however has largely been
ignored in the modern nonparametric estimation literature, where most
results only concern estimates.  Some related discussions can be found
in \citeN{gu:98b}, which also inspired the use of optimal performers
as a clone option.

\bibliographystyle{chicago}
\bibliography{root}

\end{document}